# Correction of Wall Adhesion Effects in the Centrifugal Compression of Strong Colloidal Gels

Richard Buscall[1,2,*] and Daniel R Lester[3,4]

[1] MSACT Research & Consulting, 34 Maritime Court, Haven Road, Exeter, EX2 8GP, UK.
[2] Particulate Fluids Processing Centre, Dept of Chemical & Biomolecular Engineering, University of Melbourne, Parkville, VIC 3052, Australia.
[3] CSIRO Mathematics, Informatics & Statistics, Graham Rd, Highett VIC 3190, Australia.
[4] School of Civil, Environmental & Chemical Engineering, Royal Melbourne Institute of Technology, Melbourne, VIC 3001, Australia.

*Author to whom correspondence should be addressed at r.buscall@physics.org.

## Abstract

Several methods for measuring the compressive strength of strong particulate gels are available, including the centrifuge method, whereby the strength as a function of volume-fraction is obtained parametrically from the dependence of equilibrium sediment height upon acceleration. The analysis used conventionally due to Buscall & White (1987) ignores the possibility that the particulate network might adhere to the walls of the centrifuge tube, even though many types of cohesive particulate gel can be expected to. The neglect of adhesion is justifiable when the ratio of the shear to compressive strength is small, which it can be for many systems away from the gel-point, but never very near it. The errors arising from neglect of adhesion are investigated theoretically and quantified by synthesising equilibrium sediment height versus acceleration data for various degrees of adhesion and then analysing them in the conventional manner. Approximate correction factors suggested by dimensionless analysis are then tested. The errors introduced by certain other approximations made routinely in order to render the data-inversion practicable are analysed too. For example, it shown that the error introduced by treating the acceleration vector as approximately one-dimensional is minuscule for typical centrifuge dimensions, whereas making this assumption renders the data inversion tractable.

# Introduction

Aggregated colloidal particles form compressible sediments and filter-cakes, hence the compressive strength of the particulate phase is an important property of such systems, which are sometimes called *attractive* or *cohesive* particulate gels these days. *Strongly* cohesive particulate gels, by which is meant systems where the interparticle attraction is strong enough to arrest activated processes such as coarsening, creep and delayed collapse, show self-limiting sedimentation and filtration, whereby the application of a given acceleration or load will lead to the formation of a stable sediment or filter-cake with a stable density profile. The scaled well-depth $-U_{min}/k_B T$, and hence the activation energy for particle escape and diffusion, can be very large for strongly-flocculated and coagulated colloidal particles. Hence it is found that sediments of low volume-fraction can be stable for years or even decades, as reflected by the exponential dependence of the Kramers' time upon well-depth. Under such circumstances it is meaningful to speak of equilibrium sediment heights and filter cakes thicknesses and to attempt to calculate the curve of compressive strength as a function particulate volume-fraction (φ) from curves of equilibrium filter-cake thickness (or sediment thickness) versus load (or centrifugal acceleration).

Several methods have been used to measure the compressive strength as a function of density. The most direct in concept is pressure-filtration, whereby the equilibrium filter cake volume is measured for a series of applied pressure in carefully-controlled and instrumented pressure-filters [1-8]. The analysis of this test is trivial, or at least it is provided that the possibility of adhesion of the particulate network to the side-walls of the filter can be neglected, as it can be provided that $\frac{4h\tau_w}{PD} \ll 1$, where $\tau_w$ is the adhesive shear strength, *P* the compressive strength, and *h* and *D* are the thickness and diameter of the cake respectively [9]. $\tau_w$ is taken to be the yield stress in shear at the wall and hence it can be determined independently in principle by using, say, a suitable rheometer, fitted with smooth tools made of similar material to that the walls of the filter. Alternatively, one might simply use the true yield stress of the suspension as an upper bound estimate of the adhesive strength, since it has been found in practice that whereas $\tau_y > \tau_w$, it is generally of similar order, with values of $\tau_w/\tau_y$ in the range 0.2 to 0.6 being typical of coagulated systems [8,10-13]. Another option perhaps would be to use several different filter diameters and extrapolate the results to infinite diameter, although this might not be too convenient in practice, since it requires a set of pressure filtration cells and a rather large amount of sample overall. The other three methods commonly used rely on sedimentation-equilibrium in one way or another:

(i) Measure the equilibrium sediment height in a series of batch settling columns of increasing initial sediment height under normal gravity [1,8,14-16].

(ii) Measure the equilibrium height likewise in a laboratory centrifuge fitted with either a swing-out (preferably) or a horizontal rotor for a series of accelerations [1,8, 10, 16-22].

(iii) Determine the concentration versus depth profile in a tall equilibrium sediment; tall so that there is significant compression towards the bottom [8,23-29,31,32].

The principle behind each of these methods is that the unbuoyed self-weight, $w(h) = \Delta \rho g \int_{h}^{h_{eq}} \phi(z) dz$, at any height $h$ in a sediment of total equilibrium height $h_{eq}$, is balanced by the compressive strength $P(\phi\{h\})$, where $\Delta\rho$ is the suspension inter-phase density difference, $g$ gravitational or centrifugal acceleration, $\phi(h)$ the local volume-. This assumption only holds exactly in the absence of adhesion, where the approximate 1D force balance [9,35]

$$\frac{dP}{dz} = \Delta\rho g \phi - \frac{4\tau_w}{D}, \qquad (1$$

shows wall adhesion effects can only be neglected when

$$\frac{4\tau_w[\phi(0)]}{D\Delta\rho g \phi(0)} \ll 1 \qquad (2$$

Methods (i) and (ii) are very similar in basis, clearly: the prime reason for distinguishing between them is that adhesion affects them differently as the magnitude of the acceleration $g$ is very different between centrifugation and batch settling. A secondary reason is that whereas the raw data in case (i) of $h_{eq}$ versus $h_0$ can be inverted exactly to obtain $P(\varphi)$ when adhesion is negligible, one of the corresponding pair of equations used to invert centrifuge data, that for the pressure, is now approximate. The approximation used prior to this work introduced an systematic error of ca. 5% or thereabouts, typically [18]. It is however possible to do better and render the data inversion for the centrifuge method significantly more accurate, as will be shown later in the paper.

Like pressure filtration, method (iii), concentration-profiling, is fairly direct in the absence of adhesion since integration of the concentration profile gives the pressure at any height $z$ above the base of the column, such that,

$$\Delta \rho g \int_{z}^{h_{eq}} \phi(z')dz' \to P(\phi[z]).  \tag{3}$$

It has however been shown recently [27,28] that the method is rather sensitive to any adhesion to the walls of the tube, to the point where, by using two or more columns of different diameter [27], it is possible to invert the data and determine both the compressive and the shear strength functions fairly accurately. This method is too computationally intensive to be used routinely, however a simple correction method has been recently developed [35] based upon an approximate 1D force balance, assuming the functional form of $P_y(\phi)$ is known.. The essential problem with the profiling method is that the errors coming from any adhesion accumulate up the $\phi(z)$ profile [27], as per equation (2). This can be shown clearly by consideration of (2), where the shear yield strength $\tau_w$ increases nonlinearly with $\phi$, whilst the denominator of (2) increases linearly. Hence for any tall column of modest diameter, then increasing solids volume fraction with bed depth will eventually lead to the violation of (2) and divergence of the predicted compressive yield strength somewhere down the column. Thus, the apparent or uncorrected compressive strength can appear to diverge well below the jamming limit (sometimes called point-J) when adhesion is significant in this sense [27]. The basic problem with the profiling method then is, that in order to determine $P(\varphi)$ over a significant range of concentration, the column needs to be tall in order to obtain significant compression near the bottom. This however means working at large 4 $h_{eq}/D$ which just amplifies errors coming from adhesion. Method (i), whereby the sediment is compressed more and more by increasing $h_0$ must can suffer significantly from this problem as gravitational acceleration $g$ in (2) is relatively small, order 10 [m/s$^2$]. Conversely, method (ii), centrifugation, involves much larger acceleration, where $g=\omega^2 R$, can be varied over the range $10^2$-$10^6$ [m s$^{-2}$], or more, depending upon the centrifuge. Here ω is the angular rotation rate and $R$ distance from the axis of rotation to the bottom of the centrifuge tube. Furthermore, whilst increasing initial height $h_0$ is required in sedimentation to achieve greater compressive stress in batch settling, greater pressures in centrifugation are achieved by increasing the rotation rate $\omega$, and hence acceleration $g$. Hence any errors coming from the neglect of adhesion now decrease going up the $P(\varphi)$ curve, simply because the sediment is shrinking and hence the area in contact with the wall reduces. The same is true for pressure-filtration. The other significant advantage of the centrifuge method is its large dynamic range, since four or even five decades of acceleration and thus strength can be covered given a suitable centrifuge or set of centrifuges. Its main disadvantage is that it can be difficult to get close to the gel-point since control of speed becomes a problem at very low speeds. It is likewise difficult to get close to the gel-point with pressure-filtration method because of the problem of applying and controlling small pressures.

A word needs to be said about adhesion versus cohesion perhaps. There are at least a dozen different types of interparticle attraction, or mechanisms of flocculation, known (e.g Van der Waals, depletion, liquid-bridging, incipient flocculation, charge-patch, bridging, and so on [30]). Many of these are indiscriminate inasmuch that they can produce an attraction between particles and any bounding surface too, even if the strength of attraction might depend upon the nature of the material in some cases (cf. Van der Waals), if not others (e.g. depletion). The Derjaguin approximation [30] suggests that the particle-wall force should be ca. *twice* the particle-particle force for spheres, everything else being equal. This could be taken to imply that the adhesive shear strength could perhaps be higher than the bulk shear strength, everything else being equal, although this has never been observed in experimentally to the best of our knowledge [8,10-13]. The two can be measured rheometrically by determining the shear yield stress twice, first using smooth shearing surfaces and then again with suitably roughened tools. An alternative is to use a vane tool and two smooth outer cylinders, one to give a narrow gap, such that yield occurs prematurely at the outer wall, and the second to give a wide gap such that yield then occurs at the vane [13]. Such data as there is in the literature suggests that $0.2 \leq \tau_w/\tau_y \leq 0.7$ is typical, supposing that the limited published data is reasonably representative [8,10-13].

## Constitutive relationships and modelling

Quantitative constituitive relationship were needed in order to generate synthetic sedimentation-equilibrium data. More specifically, the continuum model described in detail by Lester et al. [27] requires the compressive yield stress and wall shear functions $P_y(\varphi)$ and $\tau_w(\varphi)$ as input. Following Lester et al. [27], the functional form used for $P_y(\varphi)$ is

$$P(\phi) = k\left[\left(\frac{\phi}{\phi_g}\right)^n - 1\right]; \quad \phi > \phi_g \qquad (4$$

This form captures the experimentally-observed rapid increase in compressive strength at solids concentrations above gel-point, along with the progression towards power-law behaviour seen typically at higher concentrations. The power-law index *n* has been found to be ca. 4 typically for electrolyte-flocculated systems [8,10, 16-22,35 ] and somewhat higher for (e.g.) strongly polymer-flocculated calcium carbonate, as used in [27]. Numerical data was thus generated for both *n* = 4 and *n* = 5.

Again, following Lester et al. [27] and guided by experimental data [11], it was assumed that $\tau_w(\phi) \leq \tau_y(\phi) = \gamma_c G(\phi)$ and, in turn, that the linear shear modulus $G(\varphi) = 5/3\, K(\varphi) = 5/3\, dP(\varphi)/d\ln\varphi$. It should be emphasised that here $\gamma_c$ is merely a parameter: it is not the true shear strain, rather it is an apparent critical strain *defined* by $\gamma_c \equiv \tau_w/G$. The reason for doing so is that it is commonly found that the yield stress and modulus have similar, and in some cases identical concentration dependencies [8,10-12,19-21], meaning that the apparent yield strain, so defined, varies only weakly with volume-fraction at worst, making it a convenient parameter. Hence in the simulations, the amount of wall adhesion can be varied conveniently by changing the value of the apparent critical shear strain $\gamma_c$, defined as above, The latter has been found to vary widely in practice from one material to another. For electrolyte aggregated systems values from 0.00005 to 0.025 have been reported [8,10-12,19-21] with, 0.0001 to 0.002 being more typical perhaps. Apparent critical strains of ca. 0.02 have been reported for strongly polymer flocculated suspensions and a value of ~ 0.01 has been found enough to give strong wall effects in gravity-settling even in quite large diameter tubes [27]. From considerations of the nature of colloidal inter-particle forces the critical strain might be expected to vary inversely with particle size but only weakly if at all with volume-fraction. Published data are reasonably consistent with these notions, so far as the data go [8,10,11,21]. The functional forms used for the compressive strength combine to cause the ratio of the two, $\tau_w(\phi)/P_y(\phi)$ to decay rapidly from unity at the gel-point to an asymptotic value of $\frac{5}{3}\gamma_c$ at high volume-fraction, as shown in fig.1. The rate of decrease is such that the ratio of shear to compressive stress is approximately constant over most of the range, which is what is seen experimentally. On the other hand the ratio has to approach unity at the gel-point from theoretical considerations and, furthermore, the expected wall-adhesive effects have been observed experimentally near the gel-point [27,28]. Indeed, the constitutive forms used here were used to fit experimental data for CaCO$_3$ flocculated using three different polymeric flocculants in [27].

To generate synthetic data for analysis, sedimentation curves were generated using the functional form (4) for (apparent) critical strains $\gamma_c$ of 0.0002, 0.002 and 0.02. The gel-point and starting volume-fraction was taken to be 0.1, the initial column height $h_0$ was 0.075m and the tube diameter was 0.01m in most of the runs, latter dimensions being fairly typical of centrifuge testing.

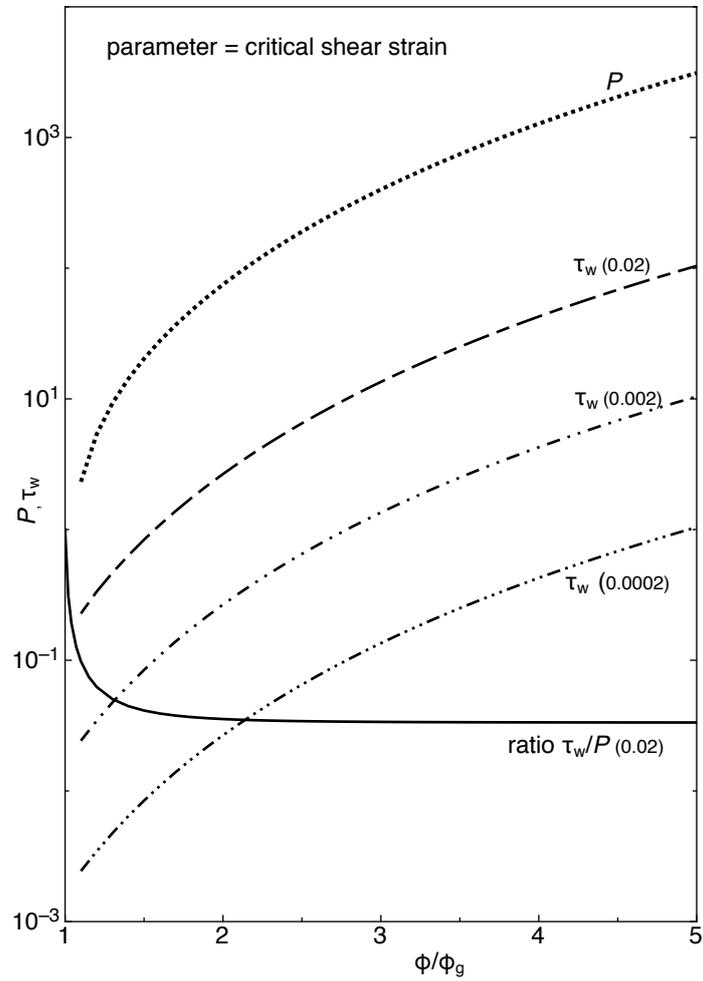

**Fig. 1** Input material functions: compressive strength and wall stress, the latter for three values of the apparent critical shear strain. The ratio of wall stress to compressive is plotted for the largest strain.

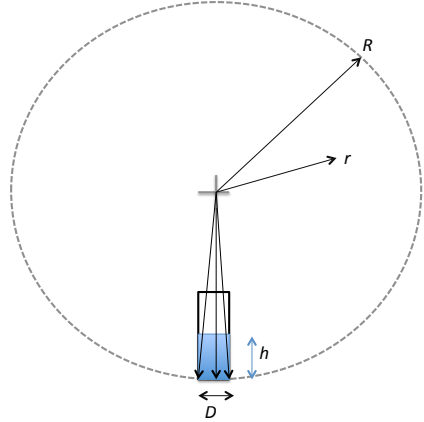

**Fig. 2** Schematic illustration of the geometry of the centrifuge test.

In any real centrifuge the acceleration $\mathbf{g} = \omega^2 \mathbf{r}$ varies with radial distance away from the axis of rotation and hence also varies down the column of suspension. It is useful to define maximum and mean values of the magnitude of the acceleration thus, (cf. Fig. 2),

$$g_{max} = \omega^2 R; \quad \bar{g} = \omega^2 (R - h_{eq}/2), \tag{5}$$

where the spatial mean represents the acceleration per unit mass for a perfectly incompressible suspension, whereas the maximum acceleration does so in the opposite limit of zero rigidity.

As we seek a simple robust means of data inversion, in most of this work the effect of wall adhesion will be explored in the uniform **g** limit, since effects of spatially-varying acceleration are typically negligible. In the case of uniform $g$ and negligible wall adhesion effects, the equilibrium curve of $h$ versus $g$ can be inverted exactly using the simple formulae [18],

$$P_y(\phi_b) = \Delta \rho g h_0 \phi_0; \quad \phi_b = \frac{h_0 \phi_0}{h_{eq} + dh_{eq}/d\ln g} \tag{6}$$

It can be seen that in this case the compressive strength then depends parametrically only upon $g$ plus three known constants, which means that any errors coming from the neglect of wall adhesion can only affect the apparent concentration at the tube bottom $\phi_b$.

In the case of radially varying $g$ the data can only be inverted approximately even when adhesion is absent or ignored [18]. The approximation suggested by Buscall & White [18] which has been used by most workers to date reads,

$$P(\phi_b) = \Delta \rho \overline{g} \phi_0 h_0$$

(7)

$$\phi_b = \frac{\phi_0 h_0 \left[1 - \frac{1}{2R}\left(h_{eq} + \frac{dh_{eq}}{d\ln g_{max}}\right)\right]}{\left[\left(h_{eq} + \frac{dh_{eq}}{d\ln g_{max}}\right)\left(1 - \frac{h_{eq}}{R}\right) + \frac{h_{eq}^2}{2R}\right]}$$

This approximation was derived by acknowledging the radial increase of $g$ down the tube axis, but by supposing otherwise that the $g$ vector field within the tube is parallel to the column walls, and equal in magnitude at any height $h$. This "parallel-g" approximation, valid for $D/2\pi R \ll 1$, simplifies the analysis of the test very considerably by rendering the analysis 1-dimensional. $D/2\pi R$ is of order 0.01 in a typical benchtop centrifuge and the error introduced by the use of the parallel-g approximation is be expected from the geometry to be of order 1-$\cos(D/R) \sim (D/R)^2$. It will be shown later via a direct comparison of simulated data for parallel and true vector **g**, that the error is indeed negligible to all intents and purposes for small $(D/2\pi R)$. Furthermore, if both $D/2\pi R \ll 1$ *and* $h/R \ll 1$, then the acceleration can be regarded as both parallel and approximately uniform (as per eqn 5), and the validity of this approximation for small but finite $(D/2\pi R)$ will be tested also.

The numerical calculations required to generate synthetic sedimentation-equilibrium curves in centrifugation, i.e. $(P_y(\phi), \tau_w(\phi)) \rightarrow (h_{eq}, \overline{g})$, can be executed efficiently via plasticity theory, although the usual problems associated with spurious slip-lines are encountered [27]. Rather, it is preferable to resolve the deviatoric stresses below the yield stress by incorporating solid-like viscoelasticity sub-yield. For materials brittle in shear (small critical strain) and for the purposes of calculating sedimentation equilibrium, the results are insensitive to the precise behaviour sub-yield and so small-strain linear elasticity suffices to regularize solutions given by plasticity theory, as has been discussed elsewhere [27]. The reader is referred to ref. [27] for details of the method employed to simulate equilibrium sedimentation curves. The values of the various constants and variables used are shown in table 1 below.

Table 1: Values of key quantities used in the simulations

| quantity | value |
|---|---|
| $\phi_g = \phi_0$ | 0.10 |
| $k$ | 5 Pa |
| $n$ | 4, 5 |
| $\Delta\rho$ | 2000 kg m$^{-3}$ |
| $D$ | 5, 10, 15 mm |
| $h_0$ | 0.075m |
| $R$ | 0.175m |

## True radial versus parallel versus uniform acceleration.

Simulated equilibrium height versus acceleration curves calculated for true vectorial **g** and constant, parallel g are plotted in fig. 2 for $D/2\pi R$ = 0.009 (D=10mm) and $h/R \sim 1/3$ and a power-law index $n$ = 4.0. It can be seen that even for $h/R$ as large as 1/3, the radial variation of acceleration makes little difference to the predicted equilibrium curve, provided that the heights from the vector simulation are plotted against the mean magnitude of the acceleration and not its maximum value. This implies that for $\frac{h}{R} \leq \frac{1}{3}$ the simpler eqn (6) can be used to invert real centrifuge data rather than (7), even though both are equally easy to use. The main difficulty in the application of either of these equations to experimental data is the calculation of the slope d$h_{eq}$/dln $g$ which needs to be evaluated as accurately as possible. Fortunately, empirical observations suggest that $h_{eq}$ as a function of ln(g) has only weak curvature (arising from power-law behaviour of $P_y(\phi)$), facilitating 1$^{st}$ order piece-wise fitting for the purpose of smoothing and differentiation. Since uniform $g$ and true vector $g$ differ only slightly, then parallel g and true vector g must differ by an even smaller amount, as shown in Fig. 3. The error introduced by the parallel-g approximation is negligible, in spite of previous criticisms regarding its use [33,34]. It is clear from the simulated data though that the parallel-g assumption is an accurate approximation for real laboratory centrifuges.

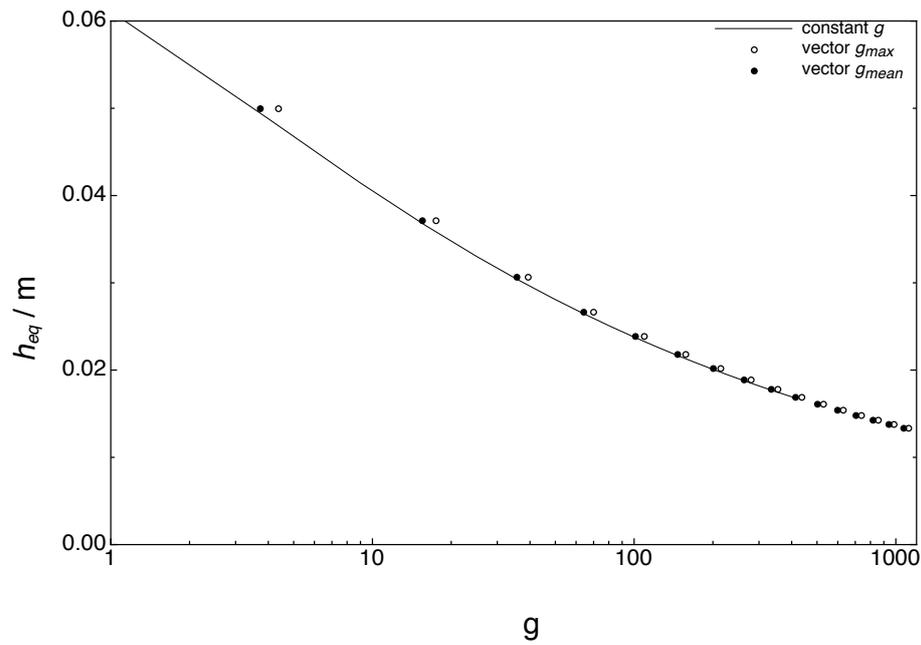

**Fig. 3** Synthesised equilibrium height versus acceleration data for constant *g equal in magnitude to the mean acceleration* $\omega^2(R-h/2)$ compared with centrifugal acceleration for $n=4$, $D=15$mm and $h_{eq}/R < 1/3$.

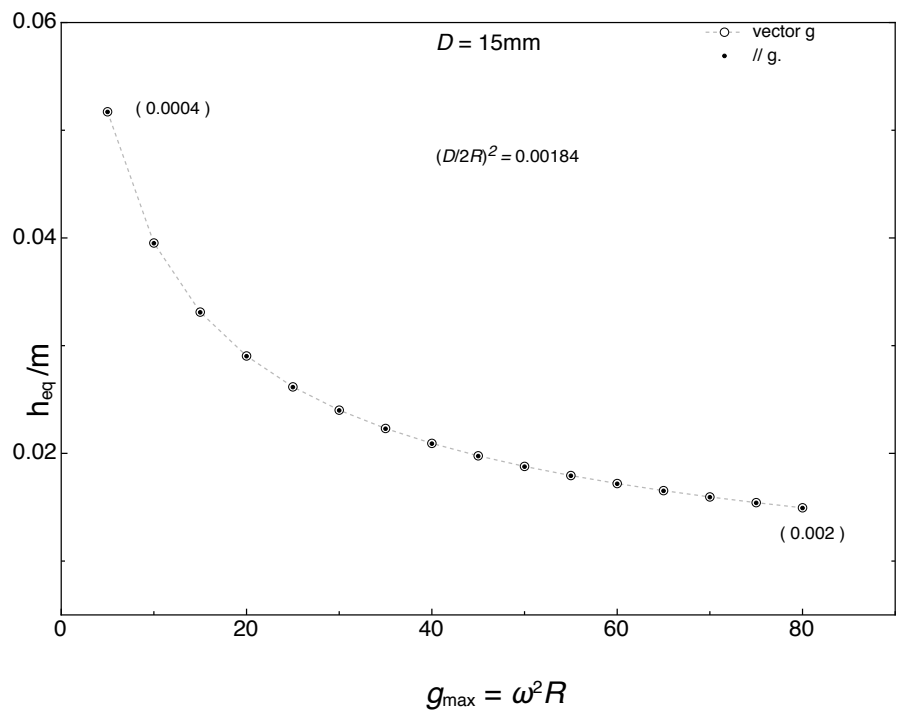

**Fig. 4** Synthesised equilibrium height versus acceleration data for vector centrifugal *g* compared with the parallel-*g* for 15mm internal dia. Tubes and *n*= 4. The figures in brackets show the fractional error of the parallel-g approximation at each end of the curve.

## Effect of adhesion

A set of synthetic equilibrium height data for both $n=5$ and $n=4$ is shown in fig. 5. Whilst the effect of increasing the critical strain from 0.0002 to 0.002 is barely discernible, that of going from 0.002 to 0.02 is more significant. For the model suspensions used here the corresponding yield stress ratio, $\tau_w(\phi)/P_y(\phi)$, decreases rapidly from ca. unity at the gel-point toward the asymptotic value of $5/3\gamma_c$ at high concentrations. A small stress ratio does not necessarily equate to a small effect though since wall effects are amplified by a geometric factor of $4h_{eq}/D$ [9, 27].

The synthetic data were inverted using (6), and some sample output is plotted in fig. 6. The large initial error introduced by neglect or ignorance of adhesion decreases rapidly with increasing volume-fraction. This decrease has two components, the first of which is the decrease of the ratio $4\tau_w(\phi(0))/D\Delta\rho g\phi(0)$ with increasing $g$, the second of which comes from the constitutive behaviour where the wall-adhesion to compressive strength ratio decreasing rapidly with volume-fraction. This behaviour is consistent with various experiments [11, loc. Cit.] and consistent too with the notion that cohesive particulate networks are short in shear but exhibit ratchet poroelastic and strain-hardening in compression [11,31,32,36,37]. Similar results for *n*= 4 are shown in fig. 6. The same pattern can be seen, large errors due neglect of adhesion for a critical strain of 0.02 near the gel-point, but with convergence on the correct result at high concentration.

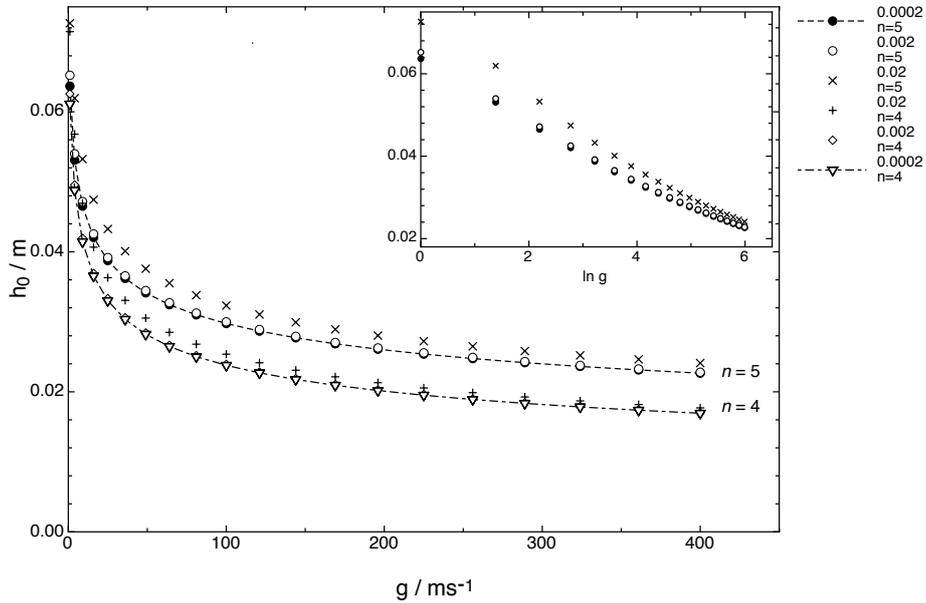

**Fig. 5** Synthesised equilibrium height versus acceleration data for constant $g$, 10mm internal diameter tubes, two values of the power-law index and for three different levels of wall adhesion as parameterised by the critical shear strain. The effect of adhesion is just discernible at 0.002 and significant at 0.02. The result for zero shear strain (not shown) is indistinguishable from the curve for 0.0002. The inset semi-log plot illustrates the slow change in the gradient term used in eqns (7) or (6).

Since rheological characterisation will often involve determination of the either or both of the adhesive or true shear yield stresses, it is pertinent to ask whether the results of applying eqns (2) or (3) be corrected, using a set of shear yield stress data measured separately using some sort of rheometer. An overall momentum balance on the sediment of the type first described by Michaels and Bolger [9] motivates the testing of corrections of the form,

$$P_{corr.}(\phi) = P_{app}(\phi) - 4\tau_w(\phi)\frac{h_0}{D}f(h_{eq})  \qquad (8$$

wherein $f(h_{eq})$ represents an approximate or average way of accounting for the effect of the consolidation of the sediment on the total adhesion. Thus, the following simple forms are suggested as likely over-corrections,

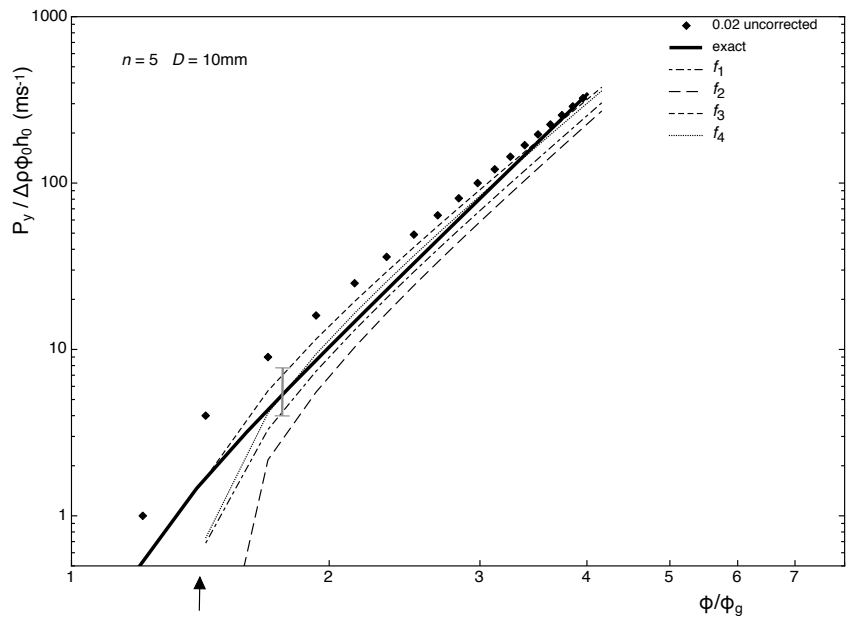

**Fig. 6** Apparent compressive strength for critical adhesive shear strain = 0.02 obtained by applying eqns 3 to the data in fig. 5 (points) compared with the true curve (heavy line). Also shown are the results of applying the suggested under and over corrections $f_1$ to $f_4$ to the points. Guessed corrections of this type cause the result to become negative below ca. 1.3 times the gel-point. Below that only iteration of the data via retrodiction and correction will do.

$$f_1(h_{eq}) = \frac{h_{eq}}{h_0} \quad \text{or,} \quad f_2(h_{eq}) = \frac{\phi_0}{\phi} \;, \tag{9}$$

whereas the following pair are likely to be under-corrections

$$f_3(h_{eq}) = \left(\frac{h_{eq}}{h_0}\right)^2 \quad \text{or,} \quad f_4(h_{eq}) = \left(\frac{\phi_0}{\phi}\right)^2 \;. \tag{10}$$

These candidate upper and lower bound corrections applied to the data for critical strain of 0.02 are compared with the true curve in figs. 6 and 7. The corrected curves bracket the true curve except near the gel-point where no simple correction term is ever going to eliminate the errors coming from unacknowledged or unknown adhesion, since it contributes hugely near the gel-point. The difficult region looks to be $\phi/\phi_g \leq 1.3$ where iteration may be required order to recover the true curve, either that or go to wider tubes, the discriminator though is the factor $\dfrac{4\tau_w h_0}{P_{y,\text{app.}} D}$. If the latter is significantly less than unity, then the lower-bound correction $f_4$ should be good, as the errors are then always fractional.

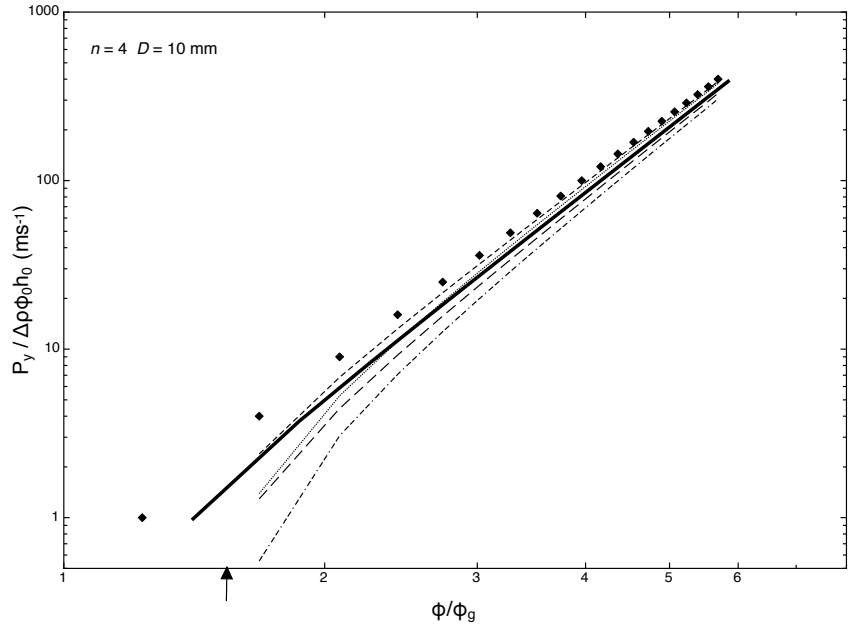

**Fig. 7** As for fig. 6 but for a power-law index of *n* = 4.  Reducing the power-law from 5 to 4 reduces the errors coming from adhesion a little, but otherwise the picture is much the same.

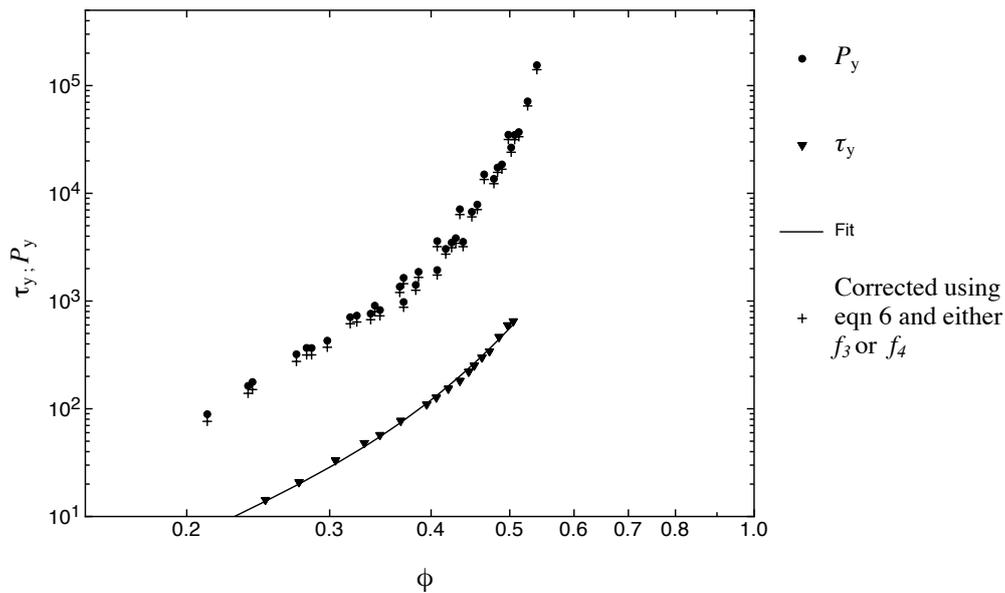

**Fig. 8** Data of Zhou et al. (replotted from ref. 10, see also ref. 8, fig. 3) and corrected according to eqn 6 with corrections $f_1$ and $f_3$ from eqns 9 and 10 using the fit (shown) to the shear data.

A data set taken from the literature is re-plotted in fig. 8, together with the lower-bound corrections calculated from the shear yield stress data shown. The latter were measured using a cruciform cross-sectioned vane and hence they are true yield stress values, and hence over-estimates of the adhesive or wall stress needed, which, in [8] is estimated as being ca. 30% of the true value for this material from a comparison of rheometric data taken from the literature [10, 21, 23]. This almost certainly means that the corrections are systematically too large, not that it would seem to matter too much in this case as they rather small anyway, since the yield stress ratio is > 10.

What if one does not have a means of generating independent shear-stress data? Or does not wish to undertake the workload (the centrifuge generates the complete curve, whereas for the shear rheometry one needs a whole series of samples spanning the concentration range and to be sure that the structure is preparation or history independent). It is fairly obvious that an alternative would be to use several values of column height and tube diameter and extrapolate each point on the sedimentation curve to $h_0/D \to 0$. It should be possible in principle to determine both strength functions by this means too, although the analysis could be rather tedious perhaps. The real problem with this method though is that one can only realistically vary $h_0/D$ by a factor of about 5 even in the most accommodating of centrifuges whereas a decade or more would probably be needed to extract reliable values for the shear strength. On the other hand, varying $h_0/D$ by just two or three can and has been used to confirm that adhesion is negligible for various materials, including coagulated latex [16-20].

The simulated effect of diameter is illustrated in fig. 9 for the most adhesive case studied here, i.e. critical strain = 0.02 and for three different values of tube diameter at fixed $h_0$ = 0.075m. The tube diameters were 5, 10, 15mm, but also shown is the datum for infinite diameter at each acceleration. The effect of diameter decreases with increasing acceleration as expected and for the two higher accelerations, and hence well away from the gel-point, the extrapolation to $1/D = 0$ is sensibly linear, whereas at g =5 it is curved away from the abscissa somewhat, implying that the use of three narrow centrifuge tubes might not be quite good enough for the most adhesive cases near the gel-point. It is always possible though to augment centrifuge data with gravity-settling data of course, where it is much easier to cover a wide a range of $h_0/D$, unless perhaps the amount of material available is very limited for some reason.

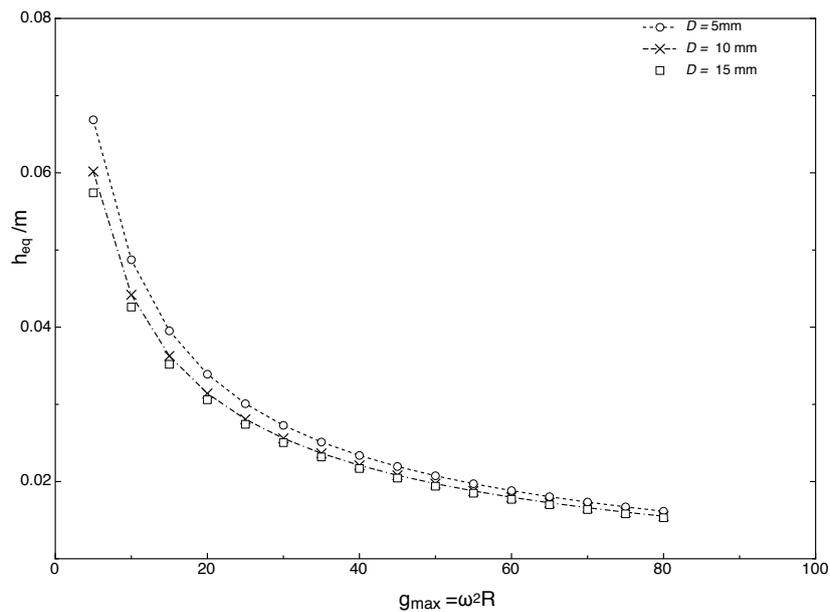

**Fig. 9** Simulated height data for a critical strain of 0.02 for three tube diameters.

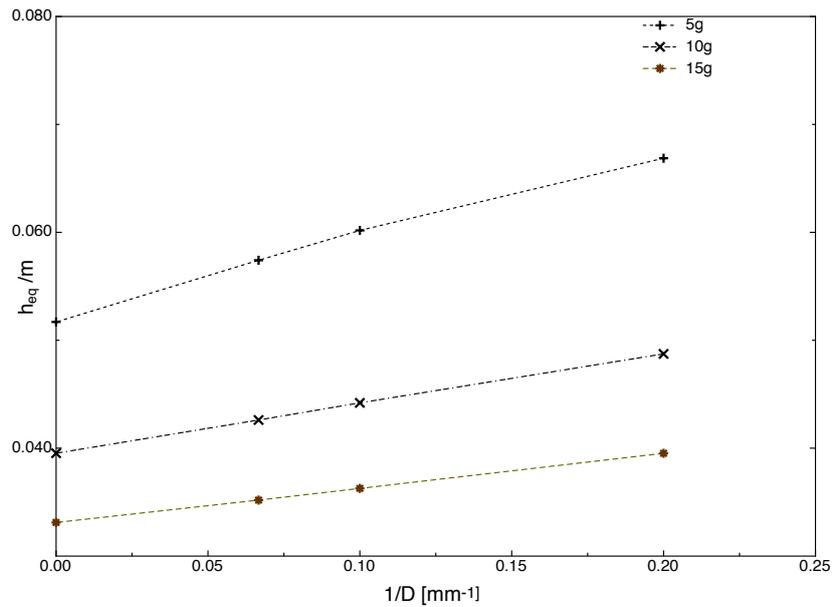

**Fig. 10** Extrapolation of simulated height data for a critical strain of 0.02 to infinite tube diameter for three accelerations measured in units of normal gravity g. Here the points for infinite *D* are known. At the lowest g, near the gel-point the plot is slightly convex to the abscissa implying that several tube diameters would be needed in practice for a good extrapolation.

## Concluding remarks

The approximate equations used conventionally for estimating the compressive strength curve for cohesive particulate gels from raw centrifuge data ignore the possibility of adhesion to the walls and furthermore are based on the assumption that the acceleration vector field inside the tube is everywhere parallel to the tube axis. The validity of these two assumptions has been explored and tested using simulated test data, as have some simple corrections for adhesion suggested by dimensional analysis. Corrections to the "parallel g" assumption are not needed as this approximation turns out produce negligible error.

Adhesion was parameterised in terms of a critical shear strain for shear yielding at the wall. This was done simply for convenience because the ratio of the shear yield stress to compressive strength itself is expected in practice to decrease rapidly with increasing volume-fraction near the gel-point, before becoming more or less constant, whereas the critical shear strain has been found to be at worst only weakly concentration-dependent. Critical shear strains are known to vary from one material to another and with particle size, hence critical strains of 0, 0.0002, 0.002 and 0.02 were used in order to cover the range reported in the literature, where values of << 0.01 have been reported for many electrolyte coagulated or flocculated suspensions, including polystyrene latex and a range of minerals, whereas values of order 0.02 have been reported for one electrolyte flocculated mineral system [27]. Aside from that study, critical strain values of order 0.01 or more have only been reported for mineral suspensions strongly-flocculated using high polymers and for water-treatment sludges. A particular form for the dependence of the shear and compressive strength on volume-fraction had to be assumed and this again was motivated by published experimental data. The dependence of the resulting adhesive to compressive strength ratio on volume-fraction and critical strain on volume-fraction and critical strain is plotted in Fig 1 in order to illustrate the parameter space covered. Small strength ratios alone do not necessarily mean a negligible effect of adhesion in the centrifuge since the relative effect of adhesion is given by the ratio $4\tau_w(\phi(0))/D\Delta\rho g\phi(0)$, hence experimental parameters such as $g$ and $D$ play a significant role.

The simulations herein suggest that adhesion can be usually neglected in centrifuge testing for volume-fractions well above the gel-point, by which is meant $> 3\phi_g$, and otherwise where the apparent critical shear strain $<< 0.01$. Otherwise, the following tactics might be considered:

1. The adhesive shear strength could be measured independently in a rheometer fitted with appropriate tools and a correction applied using (6) or (7) and (8). The simple corrections suggested here will always fail very near the gel-point though.

2. Make measurements using several different values of the ratio of initial height $h_0$ to tube diameter $D$ and extrapolate to $h_0/D$. This method may or may not be practicable, depending upon the design of the centrifuge. The range of $h_0/D$ covered should be at least three-fold and preferably five-fold, as should the number of points, although it rather depends upon the strength of the effect. Fewer points would needed to demonstrate that adhesion is insignificant, where it is, than would be needed to perform a good extrapolation, where it is not.

3. Complement the centrifuge data with data from gravity-settling for a set of column chosen so as to vary $h_0/D$ by a factor of at least five and better ten, with $h_0$ and $D$ each varied by at least three. One could also complement centrifuge data with that coming from an instrumented pressure-filter, in principle, since the construction of these is usually such that $h_0/D$ is small, except that pressure-filtration does not normally work near the gel-point because of the difficulty in applying the necessary small pressures. This method has however been used to validate and complement the centrifuge method at higher volume-fraction [8, 10, 21, 23].

The centrifuge method has been used recently to examine the effect of temperature on the stability of incipiently flocculated polystyrene latices [36], but without taking the possibility of adhesion into account. It has been used also to measure the compressibility of microgels, albeit using inclined rather than horizontal or swing-out rotors which adds complication to the analysis. Wall shear-stresses were ignored, which could well be reasonable for microgels in good solvents, perhaps [37].

and D. J. Durian, Phys. Rev. E **82**, 041403 (2010).

**Note:** One of us is sometimes asked what 'MSACT' in 'MSACT Research & Consulting'. It is short for 'Mud, Sludge and Custard Tamed', although this is more of an ambition than an claim.

21